\documentclass{aastex631}
\shorttitle{Satellite Galaxies in the TNG100
Simulation}
\shortauthors{McDonough \& Brainerd}
\graphicspath{{./}{figures/}}

\begin{document}

\title{The Distribution of Satellite Galaxies in the TNG100 Simulation}

\correspondingauthor{Tereasa Brainerd}
\email{brainerd@bu.edu}

\author[0000-0001-6928-4345]{Bryanne McDonough}
\affiliation{Boston University, Department of Astronomy
and Institute for Astrophysical Research, Boston, MA 02215}

\author[0000-0001-7917-7623]{Tereasa G. Brainerd}
\affiliation{Boston University, Department of Astronomy
and Institute for Astrophysical Research, Boston, MA 02215}



\begin{abstract}

We investigate the spatial distribution of the satellites of isolated host galaxies in the 
TNG100 simulation. In agreement with a previous,
similar analysis of the Illustris-1 simulation,
the satellites are typically poor tracers of the mean host mass density.  Unlike 
the Illustris-1 satellites, here the spatial distribution of the complete
satellite sample is well-fitted by an NFW profile; however, the concentration 
is a factor of $\sim 2$ lower than that of the mean host mass density.  The spatial distribution
of the brightest 50\% and faintest 50\% of the satellites are also well-fitted
by NFW profiles, but the concentrations differ by a factor of $\sim 2$.  
When the sample is subdivided by host color and luminosity, the number density profiles for blue satellites generally
fall below the mean host mass density profiles while the number density profiles for
red satellites generally
rise above the mean host mass density profiles.  These opposite, systematic
offsets combine to yield a moderately good agreement between the 
mean mass density profile of the brightest blue hosts and the 
corresponding number density 
profile of their satellites.  Lastly, we subdivide the satellites according to the redshifts at which they joined their hosts.  From this,
we find that neither the oldest one third of the satellites nor the youngest
one third of the satellites faithfully trace the mean 
host mass density.

\end{abstract}




\section{Introduction} \label{sec:intro}

Dark matter halos in $\Lambda$CDM simulations are known to
follow a spherically-averaged characteristic density
profile, often parameterized as a Navarro, Frenk \& White (NFW; \citealt{NFW1996}) profile:
\begin{equation}
    \frac{\rho(r)}{\rho_{\rm crit}}= \frac{\delta_c}{(r/r_s)(1+r/r_s)^2} .
\end{equation}
Here $\rho_{\rm crit}$ is the critical density for closure of the universe, $\delta_c$ is a dimensionless characteristic density and $r_s$ is a scale radius,
defined as the ratio of the virial radius, $r_{200}$, and the concentration parameter, $c$: $r_s \equiv r_{200}/c$. The characteristic density, $\delta_c$, and the concentration parameter are related through
\begin{equation}
    \delta_c= \frac{200}{3} \frac{c^3}{[\rm{ln}(1+c) - c/(1+c)]}.
\end{equation}

The NFW profile was first identified in N-body (i.e., dark matter only) simulations. While the implementation of baryonic physics in $\Lambda$CDM simulations results in dark matter halos that, in three dimensions, are rounder and more oblate than the halos in N-body simulations (see, e.g., \citealt{Chua}), the spherically-averaged density profiles of the dark matter halos in these cosmological hydrodynamical simulations still follow NFW density profiles at all but the smallest radii (see, e.g., \citealt{Brainerd2018}).

The shapes of the density profiles of dark matter
halos in the observed universe, and whether or not they
follow NFW-like profiles, are
important tests of $\Lambda$CDM.
Given the invisible nature of the dark matter, some type of luminous tracer of the 
dark matter halos of observed galaxies is, of course, necessary in order
to directly constrain their mass density profiles.  Satellite galaxies, orbiting 
about bright, central ``host'' galaxies, could potentially serve as such tracers. 
However, the degree to which satellite galaxies faithfully
trace the underlying mass distributions of their host galaxies is unclear.

\cite{Budzynski2012} found that the radial number density profiles of the
satellites in observed groups and clusters of galaxies
were well-fitted by NFW profiles.  Compared to predictions from
$\Lambda$CDM, however, the concentration found by \cite{Budzynski2012} 
was a factor of $\sim 2$ lower than expected. 
In another study of observed galaxy clusters, \cite{Wang2018} compared the locations 
of the satellites to the weak lensing-derived
cluster mass distribution.  From this, \cite{Wang2018} found that the satellite number density traced the mass density well, with both the satellite number density and the 
mass density being well-fitted by an NFW profile.

In a study of the locations of the satellites of massive red host galaxies,
\cite{Tal2012} found that, at large distances from their hosts, the radial number density profile of the
satellite galaxies could be fitted by an NFW profile. Close to the central galaxy, however, \cite{Tal2012} found that the satellite number density
exceeded the expectations of an NFW profile. Similarly,
\cite{Watson2012} investigated the small-scale clustering of satellite galaxies in the Sloan Digital Sky Survey (SDSS; \citealt{York2000}) and found that, while the spatial distribution of satellites with $M_r > -19.5$ was fitted well by an NFW profile, the spatial distribution of brighter satellites had a slope that
was steeper than expected from an NFW profile. In another study of
SDSS galaxies, \cite{Piscionere2015} found that the distribution of their most luminous satellites ($M_r < -20$) showed a steep upturn at small distances from the host galaxy. 
In a study of isolated SDSS host galaxies, \cite{Wang2014} found 
that the spatial distribution of the satellites of high mass hosts
($M_\ast > 10^{11} M_\odot$)
was less concentrated than would be expected for the hosts' dark matter halos, but that
the satellites of low mass hosts followed the expected dark matter distribution.
In contrast, \cite{Guo2012} found that the spatial
distribution of the satellites of isolated SDSS host galaxies agreed with an NFW profile, with small exceptions at small radii or low luminosity.

Mixed conclusions regarding the degree to which satellite galaxies in
$\Lambda$CDM simulations trace 
the surrounding mass distribution have also been reached.
\cite{Gao2004} applied a semi-analytic galaxy formation model (SAM) to 10 high resolution resimulations of cluster-mass halos and found that the radial distribution of luminous galaxies closely followed that of the dark matter. Using a numerical hydrodynamical simulation, \cite{Nagai2005} studied
the distribution of the satellites in galaxy clusters and concluded
that the satellite distribution was less concentrated than the dark matter, particularly in the inner regions of the clusters. 
\cite{Sales2007} applied a SAM to the 
Millennium simulation (\citealt{Millennium}) and found that the spatial distribution of the satellites of bright, isolated host galaxies was well-fitted by an NFW profile that was only slightly less concentrated than the best-fitting NFW profile for the host galaxies' dark matter halos.
Using a SAM applied to the Millennium and Millennium II (\citealt{MillenniumII}) simulations, \cite{Wang2014} found that the
satellite distribution around isolated host galaxies traced the hosts' dark matter halos well. Similarly, \cite{Guo2013} found that the radial distributions of the satellites of isolated host galaxies in the Millennium and Millennium II
simulations could be
fitted by NFW profiles. However, when \cite{Guo2013} subdivided their samples by color they found that, because the simulation produced too few blue satellite galaxies 
compared to the observed universe, the shapes of the number density profiles of their
simulated and observed galaxies were no longer similar to each other. 

More recently \cite{Agustsson2018} investigated the spatial distributions of satellite galaxies that were selected from a mock redshift survey of the Millennium simulation. Using satellites that were obtained using redshift space proximity criteria, \cite{Agustsson2018} found that the locations of the satellites of isolated red host galaxies traced the hosts' dark matter halos well.  In the case of isolated blue host galaxies, however, \cite{Agustsson2018} found that the concentration of the satellite distribution was roughly twice that of the hosts' dark matter halos.  In addition, the locations of satellite galaxies in the Illustris-1 simulation \citep{Nelson2015,Vogelsberger2014} were investigated by \cite{Ye2017} and \cite{Brainerd2018}. \cite{Ye2017} focused on halos with virial masses in the range $10^{12} h^{-1} M_{\odot}<M_{200} < 10^{14} h^{-1} M_{\odot}$, while \cite{Brainerd2018} focused on isolated host galaxies with median virial masses $\sim 10^{12} M_\odot$.  \cite{Ye2017} found that the locations of the satellite galaxies traced the dark matter halos well at large radii, but for radii $\lesssim 0.4r_{200}$ the dark matter distribution was traced well by only those
satellites with a high satellite-to-host mass ratio. \cite{Brainerd2018} found that the satellite number density profile was inconsistent with an NFW profile and, for
small host-satellite separations, the satellite number density profile increased steeply.

The disparate conclusions that have been reached regarding the relationship
between satellite number density profiles and host mass density profiles
may indicate that 
the sample selection criteria and/or the models in the simulations
influence the degree to which satellite galaxies are found to trace the local mass distribution.
Here, we revisit the question of whether or not the locations of satellite galaxies
in a $\Lambda$CDM simulation faithfully trace the dark matter distribution that surrounds
isolated host galaxies.  Using a hydrodynamical simulation that reproduces the 
red-blue color bimodality of galaxies in the observed universe, we compute the locations
of the satellites as a function of the colors and luminosities of the host and satellite galaxies.
The paper is organized as follows. The selection of the host galaxies and their satellites, as
well as the properties of the sample, are described in
\S \ref{sec2}. Normalized satellite number density profiles are
compared to normalized mass density profiles of the host galaxies in
\S \ref{sec3}.  A summary and discussion of the results
is presented in \S \ref{disc} and our main conclusions are
presented in \S \ref{conc}.

\begin{figure*}
\begin{center}
	\includegraphics[scale=0.95]{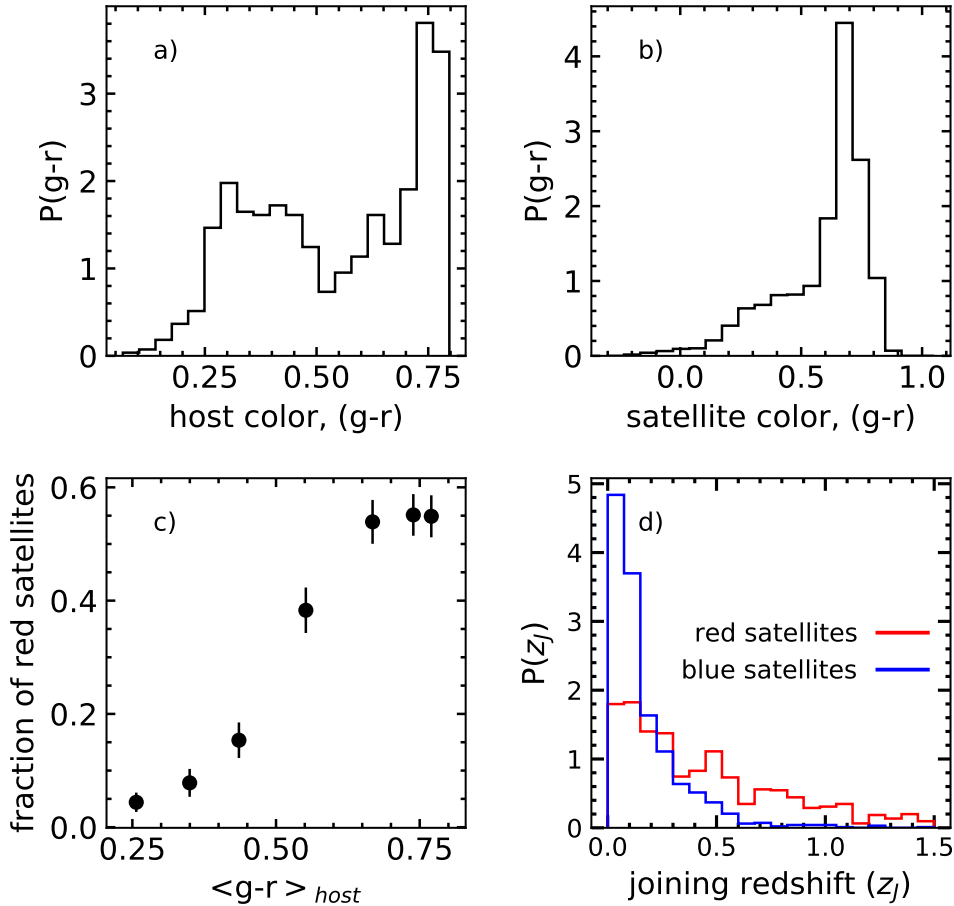}
    \caption{Properties of the host-satellite sample.  a) Probability
    distribution of host $(g-r)$ colors.
    b) Probability distribution of satellite $(g-r)$ colors.
    c) Fraction of satellites within the virial radius that
    are red in color, computed as a function of host color, d) Probability distribution
    of the redshifts at which the satellites joined their hosts' halos, computed
    separately for red and blue satellites.  Probability distributions in
    panel d) have been truncated at $z_j = 1.5$ to improve legibility.
    Error bars for the data points in panel c) were computed using
    1,000 bootstrap resamplings of the data.}
    \label{hosts_and_sats}
\end{center}
\end{figure*}

\newpage

\section{Host-Satellite Sample} \label{sec2}

Our host-satellite sample is drawn from the $z = 0$ snapshot of the
TNG100-1 simulation.  The IllustrisTNG project \citep{Nelson2018,2018MNRAS.475..648P,2018MNRAS.475..676S,2018MNRAS.477.1206N,2018MNRAS.480.5113M,Nelson2019a} is a suite of $\Lambda$CDM magnetohydrodynamical simulations, all of which are publicly available. The simulations
adopted a \citet{Planck2016} cosmology with the following parameter values:
$\Omega_{\Lambda,0}=0.6911$, $\Omega_{m,0}=0.3089$, $\Omega_{b,0}=0.0486$, $\sigma_8=0.8159$, $n_s=0.9667$, and $h=0.6774$.  
Here, the TNG100 simulation was chosen 
because it offers good resolution of relatively faint
satellite galaxies ($M_r \sim -14.5$).  In addition, the number of dark matter
particles ($1820^3$), the number of gas particles at the start of the simulation
($1820^3$), the
co-moving volume of the simulation ($75^3 h^{-3}~\rm{Mpc}^3$), and the initial conditions
are identical to those of 
the Illustris-1 simulation.  This makes
comparison between our results and those of \citet{Brainerd2018} straightforward.  Excluding
subhalos that are flagged as being non-cosmological in origin, there are $\sim 60,000$ ($M_r < -14.5$) luminous
galaxies in the TNG100 simulation at $z = 0$.

Key differences between the original Illustris simulations and the IllustrisTNG
simulations are the ways in which AGN feedback, galactic winds, and the growth of supermassive black holes were treated (see \citealt{Pillepich2018,Weinberger2017}).  
These changes successfully addressed a shortcoming of the original Illustris simulations: the lack of a clear red-blue color bimodality for present-day galaxies.  Unlike
Illustris-1, the TNG100 simulation shows a
red-blue color bimodality for present day galaxies that is comparable to the color distribution
of low-redshift galaxies in the observed universe.  
For rest frame Sloan $(g-r)$ color, the division between ``red'' and ``blue'' galaxies occurs at $(g-r) = 0.65$ for TNG100 galaxies at redshift $z = 0$ (see \citealt{Nelson2018}).

In order to compare our results from TNG100 to those obtained from Illustris-1, we adopt the same 
3-D selection criteria as \citet{Brainerd2018}.  Each host galaxy must have an absolute
magnitude brighter than $-20.5$ in the $r$-band, be located at the
center of its friends-of-friends halo and 
be at least 2
magnitudes brighter than any other galaxy within a distance
of $1 h^{-1}$~Mpc.  The last of these 
criteria insures that the host galaxies are ``isolated'' and, in general, the criteria we 
have adopted will select host-satellite systems that are more isolated than the Milky Way or M31.  Although isolation insures that the hosts are at the centers of the potentials and that the satellites do not experience significant gravitational effects due to other nearby, large galaxies, isolation also results in the selection of bright central galaxies that have relatively few neighboring galaxies compared to non-isolated bright central galaxies.  
For the bright, central galaxies that satisfy the above criteria, all galaxies with $M_r < -14.5$
that are within 
$2r_{200}$ of the bright galaxy are defined to be satellites.  To ultimately qualify as being a ``host'' in our sample, the bright central galaxy must have at least 1 satellite within the virial radius.

The galaxy magnitudes that we use here are the sum of the intrinsic emission of all bound star particles, as described in \cite{2018MNRAS.475..648P}, and the effects of dust obscuration are not included. Following
\citet{Nelson2018}, we use the $(g-r)$ colors of the hosts 
and satellites to split our sample into ``red'' and ``blue'' galaxies, where red galaxies
are those with $(g-r) \ge 0.65$ and blue galaxies are
those with $(g-r) < 0.65$. We have verified that our results below are not overly sensitive to the precise value of $(g-r)$ color that is used to separate the galaxies according to color.

Table~1 lists various properties of the host galaxies as a function of host color.  In order, these are:
number of hosts, median $r-$band absolute magnitude, median number of satellites within the virial radius, maximum number of satellites in a system,
median stellar mass, median virial mass, minimum virial mass, maximum virial mass, median virial radius, minimum virial radius, and maximum virial radius. From Table~1, although red and blue hosts are similarly bright, red hosts have stellar masses and virial masses that are systematically larger than those of blue hosts.  In addition, red hosts typically have more satellites within their virial radii than do blue hosts.  From Table~1, the selection criteria result the inclusion of a handful of group- and cluster-sized objects.  This is also true of the host-satellite samples in
\cite{Brainerd2018} and \cite{Sales2007} (which were obtained using the same criteria as we adopt here), and does not affect our conclusions below.  \cite{Brainerd2018} investigated the effects of inclusion of these larger systems and found that they did not significantly affect the conclusions regarding the degree to which the satellite distributions differed from the host mass distributions.

\begin{deluxetable*}{ccc}
\tablenum{1}
\tablecaption{Host galaxy properties}
\tablewidth{0pt}
\tablehead{property & red hosts & blue hosts}
\startdata
number & 286 & 461 \\
$M_r^{\rm med}$ & -21.7 & -21.5 \\
$N_{\rm sat}^{\rm med}$ & 2 & 1 \\
$N_{\rm sat}^{\rm max}$ & 145 & 230 \\
$M_\ast^{\rm med}$ & $5.65\times 10^{10} M_\odot$ & $2.39\times 10^{10} M_\odot$ \\
$M_{200}^{\rm med}$ & $2.79 \times 10^{12} M_\odot$ & $1.05\times 10^{12} M_\odot$ \\
$M_{200}^{\rm min}$ & $1.20 \times 10^{12} M_\odot$ & $5.17 \times 10^{11} M_\odot$ \\
$M_{200}^{\rm max}$ & $2.13 \times 10^{14} M_\odot$ & $1.71 \times 10^{14} M_\odot$ \\
$R_{200}^{\rm med}$ & $297 \, \rm kpc$ & $214 \, \rm kpc$ \\
$R_{200}^{\rm min}$ & $197 \, \rm kpc$ & $149 \, \rm kpc$\\
$R_{200}^{\rm max}$ & $1,260 \, \rm kpc$ & $1,170 \, \rm kpc$\\
\enddata
\end{deluxetable*}

The top panels of Figure~\ref{hosts_and_sats} show the $(g-r)$ color distributions 
of the host and satellite galaxies, from which it is clear that the majority of
the hosts are blue and the majority of the satellites are red.  Figure~\ref{hosts_and_sats}c)
shows the fraction of satellites inside the virial radius that are red in color, computed as a function of host color.  From Figure~\ref{hosts_and_sats}c), a clear color-color correlation
exists for the host galaxies and their satellites, with blue hosts having a smaller
fraction of satellites that are red in color and $\gtrsim 50$\% of the satellites 
of red hosts being red in color.

Figure~\ref{hosts_and_sats}d) shows probability distributions for
the redshifts at which the satellites first approached within the present-day virial
radii of their host galaxies. Throughout, we will refer to this redshift as the satellite's ``joining redshift'', $z_j$. Splashback satellites (i.e., those that enter, leave, and re-enter their host's virial radius) are counted as having ``joined'' the host upon their first entrance. Flyby satellites (i.e., those that will not end up bound to the halo) are not excluded from our satellite sample, but would necessarily have very recent joining redshifts since the satellite sample is drawn from the $z=0$ timestep of the simulation. As expected, most of the satellites that are red at the present day entered their hosts' halos earlier than the satellites that are blue at the present day. Some
satellites that are red at the present day entered their hosts' halos for the first time at redshifts $> 10$, resulting in a mean joining redshift for the red satellites of $\left< z_j \right> = 1.0$. In the case of satellites that are blue at the present
day, relatively few had entered their hosts' halos by a redshift of $\sim 1$, resulting in a mean joining redshift for the blue satellites of $\left< z_j \right> = 0.24$. The long tail on the probability distribution of $z_j$ for the red satellites gives rise to a difference of $\sim 5$~Gyr of cosmic time between the mean joining redshifts of the red and blue satellites;
hence, {\it on average} the red satellites are an ``older'' population of satellite galaxies.
However, the median joining redshifts (i.e., the redshifts at which 50\% of each type of satellite had joined their hosts' halos) are more similar than are the mean joining redshifts ($z_{j,{\rm med}} = 0.4$ for the red satellites vs.\ $z_{j,{\rm med}} = 0.11$ for the blue satellites, or a difference of only $\sim 2$~Gyr of cosmic time).

\section{Normalized Radial Density Profiles} \label{sec3}

In this section we present results for the radial mass density profiles
of the host galaxies, together with the radial number density profiles
for the surrounding satellite galaxies.  Following \cite{Sales2007} and \cite{Brainerd2018},
we normalize the density profiles to unity at the virial
radius.  We also define a dimensionless radius, $x \equiv r/r_{200}$, so that the normalized
density profile is given by $\rho(x)/\rho(x=1)$, where $\rho(x)$ is the
density profile prior to normalization.  Throughout, we compute the radial
mass density profiles for each host galaxy using all of the surrounding dark matter, gas, and
stellar particles.  We then compute the mean radial mass density profile using
either all host galaxies or particular subsets of the host galaxies.  
Error bars are computed using 1,000 bootstrap 
resamplings of the data and are omitted when they are comparable to or
smaller than the data points.

In both the observed universe and in simulation space, isolated host-satellite systems in which the hosts have masses similar to those in our
study (i.e., $M_{200} \sim 10^{12} M_\odot$) contain relatively few bright satellites when compared to systems with less isolated hosts. 
Therefore, 
satellite number density profiles are not computed directly for individual
systems.  Instead, ensemble-averaged number density profiles are computed by
stacking the host-satellite systems together (see, e.g., \citealt{Sales2007}, \citealt{Tal2012}, \citealt{Guo2013}, \citealt{Wang2014}, \citealt{Ye2017}, \citealt{Brainerd2018}).  Even in the case of galaxy
clusters, ensemble-averaged number density profiles are commonly used to assess
the degree to which the spatial distribution of the cluster members traces that of the 
dark matter distribution (see, e.g., \citealt{Gao2004}, \citealt{Budzynski2012}, \citealt{Ye2017},
\citealt{Wang2018}).

Below we adopt this standard
technique and compute ensemble-averaged
radial number density profiles for the host-satellite systems,
stacked according to various criteria.  For comparison with the 
shapes of the normalized mass
density profiles, we also normalize the satellite number density profiles
to unity at $x = 1$.
Again, error bars are computed using 1,000 bootstrap 
resamplings of the data and are omitted when they are comparable to or
smaller than the data points.

\begin{figure}
\begin{center}
	\includegraphics[scale=0.95]{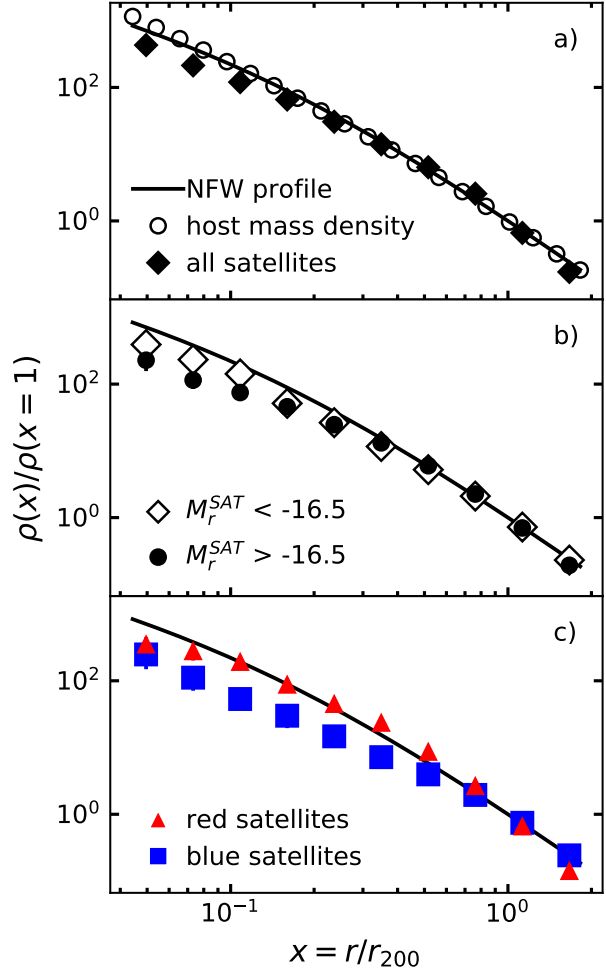}
    \caption{Normalized radial density profiles for the TNG100
    host-satellite systems. 
    Black lines: NFW profile that best fits the mean radial mass density
    of the host galaxies (concentration parameter: $c=8.9)$.
    a) Circles: mean radial mass density of the host galaxies.
    Diamonds: mean radial number density of all satellite galaxies.
    b) Diamonds: mean radial number density for satellites 
    with $M_r < -16.5$.  
    Circles: mean radial number density for satellites
    with $M_r > -16.5$. 
    c) Mean satellite radial number density, computed separately for
    red and blue satellites.}
    \label{tng_profiles}
\end{center}
\end{figure}

Circles in Figure \ref{tng_profiles}a) show the mean radial mass density of all host
galaxies. The solid black line in Figure \ref{tng_profiles}a) shows the best-fitting
NFW profile, obtained using a non-linear least squares regression.  The best-fitting NFW profile has a concentration parameter of $c = 8.9$ and is formally a good fit to the mean
radial mass density of the hosts. The mean radial number density
of all satellites is shown by the diamonds in Figure \ref{tng_profiles}a), 
from which it is clear that for 
$x \lesssim 0.2$ the normalized satellite number density profile falls below
that of the host mass density. Therefore, the satellites do not 
trace the host mass density on scales $\lesssim 20$\% of the
virial radius. In the innermost regions of the host-satellite systems, the
normalized number density profile for the satellite galaxies in
Figure \ref{tng_profiles}a) is only $\sim 40$\% of the corresponding mass density
profile of the host galaxies.
The satellite distribution in Figure \ref{tng_profiles}a) is
well-fitted by an
NFW profile with concentration parameter $c=4.8$, indicating that
the spatial distribution of the satellites is considerably less concentrated
than that of the host mass.

\begin{figure}
\begin{center}
	\includegraphics[scale=1.0]{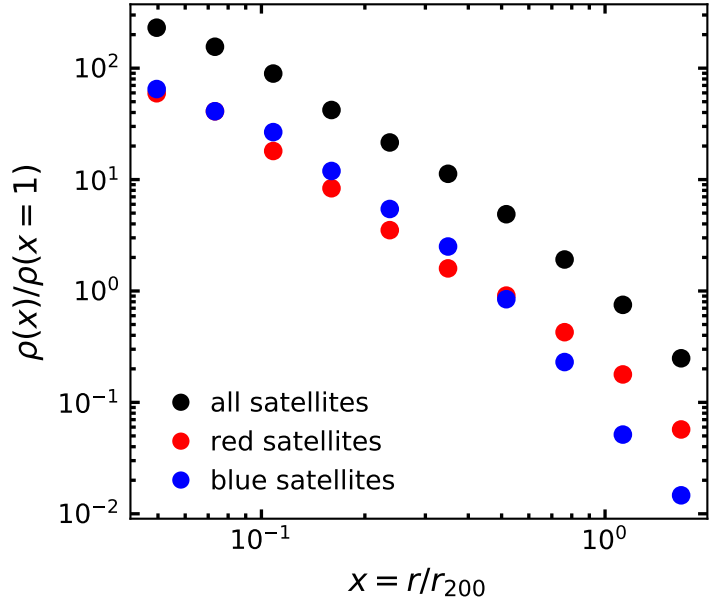}
    \caption{Contribution of the reddest one third of all
    satellite galaxies and the bluest one third of all satellite galaxies 
    to the normalized radial number density profile that was obtained
    using the complete sample of TNG100 satellite galaxies.}
    \label{red_blue_sats}
\end{center}
\end{figure}

The deviation between the shapes of the mean halo mass density profile
and the satellite number density profile is explored further Figure \ref{tng_profiles}b) and \ref{tng_profiles}c)
where we subdivide the satellite sample
by $r$-band absolute magnitude and $(g-r)$ color.  
In Figure \ref{tng_profiles}b), we divide the satellites into the
brightest 50\% of the sample and the faintest
50\% of the sample using the median satellite $r$-band
absolute magnitude, $M_r^{\rm med} = -16.5$.
From Figure \ref{tng_profiles}b),
neither the brightest nor fainest satellites 
reproduce the shape of the mean halo mass density
profile on scales $\lesssim 25$\% of the virial radius; however, the
discrepancy between the mean halo mass density profile and the satellite number
density profile is greater for the faintest satellites than it is for the 
brightest satellites.  The spatial distributions of both the brightest
and the faintest satellite galaxies are well-fitted by NFW profiles with
concentrations $c = 4.2$ and $c = 2.3$, respectively.  That is,
while both populations are distributed in a manner that is consistent
with an NFW profile, both populations are less concentrated than
the mean host mass density, and the spatial distribution of the faintest
satellites is considerably less concentrated than the spatial distribution
of the brightest satellites.  

Figure \ref{tng_profiles}c) shows the dependence of the
normalized radial number density profile on the
satellites' optical color (2366 red satellites vs.\ 1320 blue satellites).
From Figure \ref{tng_profiles}c), neither red nor blue satellites reproduce the shape of the mean halo mass density profile over all scales at which satellites are found.  In the case of the red satellites, the number density profile falls below that of the mean halo mass density profile on scales $\lesssim 10$\% of the virial radius.  In the case of the blue satellites, the number density profile falls below that of the mean halo mass density profile on scales $\lesssim 50$\% of the virial radius.  
Additionally, the spatial distribution
of the blue satellites is well-fitted by an NFW profile with concentration
$c = 1.9$; however, the spatial distribution of the red satellites is
not well-fitted by an NFW profile.

\begin{figure*}
\begin{center}
	\includegraphics[scale=1.0]{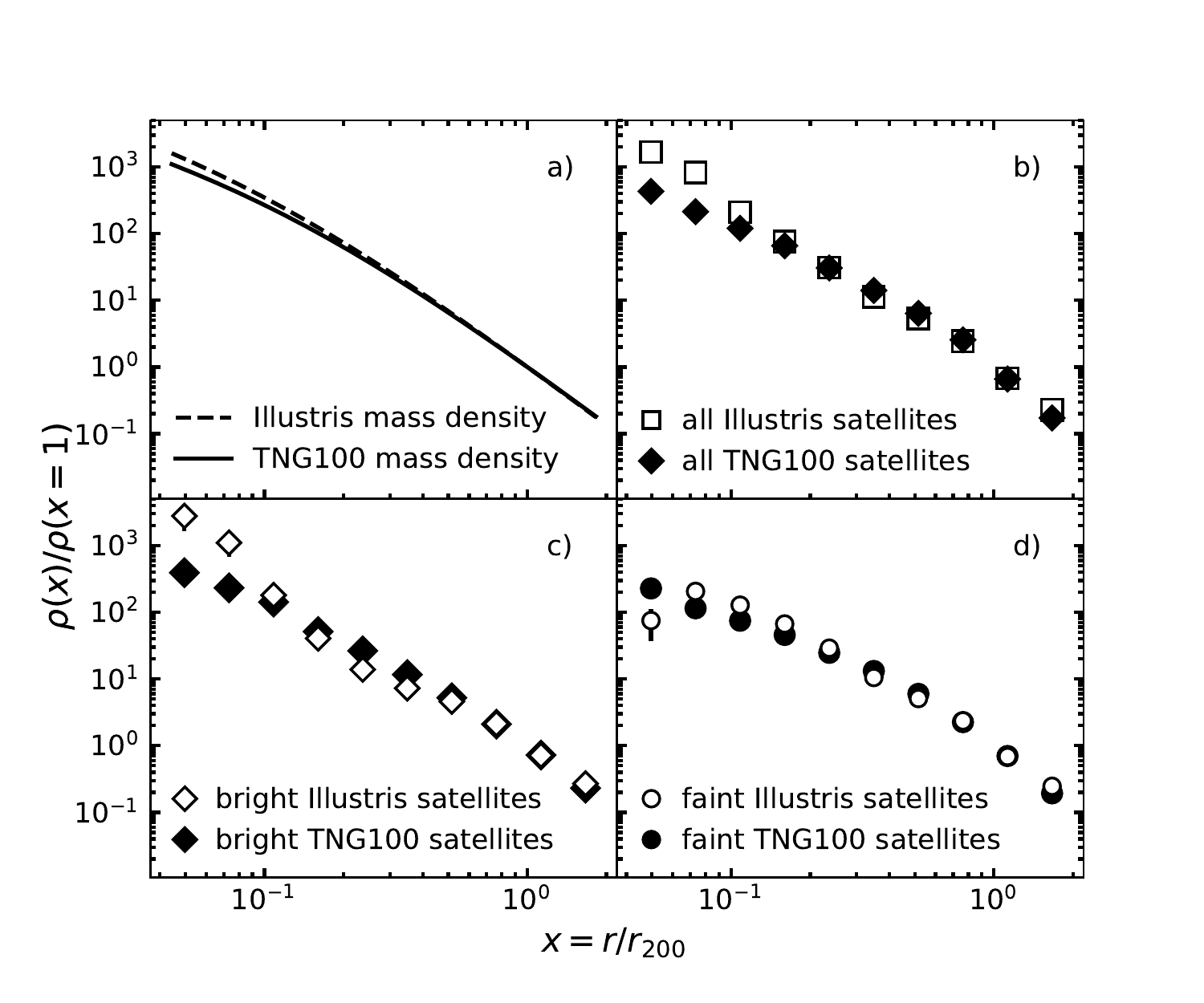}
    \caption{Comparison of TNG100 host-satellite
    systems and Illustris-1 host-satellite systems. a) Best fitting NFW profiles for the mean host mass densities.
    b) Radial number density profiles obtained using all. c) Radial number density profiles obtained using only satellites
    with $M_r < -16.5$.
    d) Radial number density profiles obtained using
    only satellites with $M_r > -16.5$.}
    \label{tng_illustris}
\end{center}    
\end{figure*}

Next, we divide the full satellite sample into three parts according to their
$(g-r)$ colors. Figure \ref{red_blue_sats} shows the individual contributions of the reddest one third and bluest one third of the complete satellite sample to the normalized number density
profile that was computed using all TNG100 satellites (i.e., 
the results shown in Figure~\ref{tng_profiles}a).  That is, in
Figure \ref{red_blue_sats} we do not normalize the number density profiles
of the reddest and bluest satellites to unity at $x=1$.  Instead, they
are normalized to the number density of all TNG100 satellites
at $x=1$. This allows us to determine what fraction of the number
density profile for all satellites is associated with the reddest satellites
vs.\ the bluest satellites.  

Colored points in Figure \ref{red_blue_sats} show
results for the reddest one third of all satellites
($(g-r) > 0.69$; red points) and the bluest one
third of all satellites ($(g-r) < 0.59$; blue points).
From this, it is clear that the distributions of the reddest and bluest
satellites contribute similar amounts to the overall
satellite number density profile at small 
host-satellite separations.  This is in stark contrast to the 
results of \cite{Sales2007}, who applied the same 
host-satellite selection
criteria to a
semi-analytic galaxy catalog that was constructed using the
Millennium simulation.  \cite{Sales2007} found that the distribution
of the reddest one third of their satellites was considerably 
more concentrated than was the
distribution of the bluest one third of their satellites. In
the case of the \cite{Sales2007} satellites, 50\% of the reddest
one third of the satellites
were found within a host-satellite separation of $0.32r_{200}$, while 50\%
of the bluest one third of the
satellites were found within a host-satellite separation
of $0.71r_{200}$.  For the TNG100 satellites
we find 50\% of the reddest one third of the satellites are within $0.49r_{200}$ and 50\% of the bluest one third of the satellites are within $0.53r_{200}$. Compared to the results of \cite{Sales2007}, the bluest and reddest TNG100 satellites are more equally distributed across the radial extent of the host. The difference in these simulated samples likely arises from the semi-analytic model used by \cite{Sales2007}, which assumed that satellites immediately lose their reservoir of star-forming gas when they are accreted by a host. This leads to a fast color transition from blue to red, resulting in blue satellites being found primarily in the outer regions of their hosts' halos. In TNG100, gas must be removed from the satellites via hydrodynamical processes on a longer timescale, resulting in blue satellites that can exist for longer times within their hosts' halos and, therefore, can be found at small host-satellite distances.

\subsection{Comparison of TNG100 and Illustris-1 Host-Satellite Systems}

As noted above, the models used for AGN feedback, galactic winds, and 
the growth of supermassive black holes in the TNG100 simulation differed from those used in the Illustris-1 simulation.  These modifications resulted in a significant change to the color distribution of present-day galaxies.  In addition, they have the potential to affect both the mean host halo concentration and the spatial distribution of the satellites.  Therefore, in Figure \ref{tng_illustris} we compare results from our TNG100 sample to results from an 
identically-selected Illusris-1 sample (e.g.,  \citealt{Brainerd2018}).  In both the Illustris-1 sample and the
TNG100 sample, the mean host mass density profiles are well-fitted by NFW profiles, and the best-fitting NFW profiles for both sets of host galaxies are shown in 
Figure~\ref{tng_illustris}a).  From \cite{Brainerd2018}, the best-fitting NFW profile 
for the mass density of the Illustris-1 hosts has a concentration parameter of $c=11.9$,
which is considerably larger than the value of $c=8.9$ that we find for the TNG100 hosts.  This is consistent with the fact that the concentration parameters of galaxies in the original Illustris suite of simulations were higher than expected (see \citealt{2018MNRAS.475..648P}) and the updated models in the TNG suite of simulations resulted in galaxies that are less concentrated and which better match observed galaxies. 

In addition to differences in the mean host mass density, the satellite distributions in the two simulations also show differences. Figure \ref{tng_illustris}b) shows satellite radial number density profiles computed using all of the satellites in each simulation. From this, it is clear that the number density profile of the Illustris-1 satellites exceeds that of the TNG100 satellites on scales $\lesssim 10$\% of the
virial radius and the discrepancy is most pronounced for satellites brighter than $M_r = -16.5$ (see Figure \ref{tng_illustris}c).  For these bright satellites, the number
density profiles from Illustris-1 and TNG100 show good agreement only on scales 
$\gtrsim 0.6r_{200}$.  On scales $\lesssim 0.1r_{200}$, the disagreement between the number density profiles for satellites with
$M_r < -16.5$ correlates with the steep increase in the number density profile found by \cite{Brainerd2018} for bright Illustris-1
satellites with small host-satellite separations.  Since the two host-satellite samples were selected in exactly the same manner, 
one of the outcomes of the improved models in the TNG suite of simulations is the elimination of the
steep increase in the satellite number density found by \cite{Brainerd2018} at small host-satellite separations.

\begin{figure}
\begin{center}
	\includegraphics[scale=0.95]{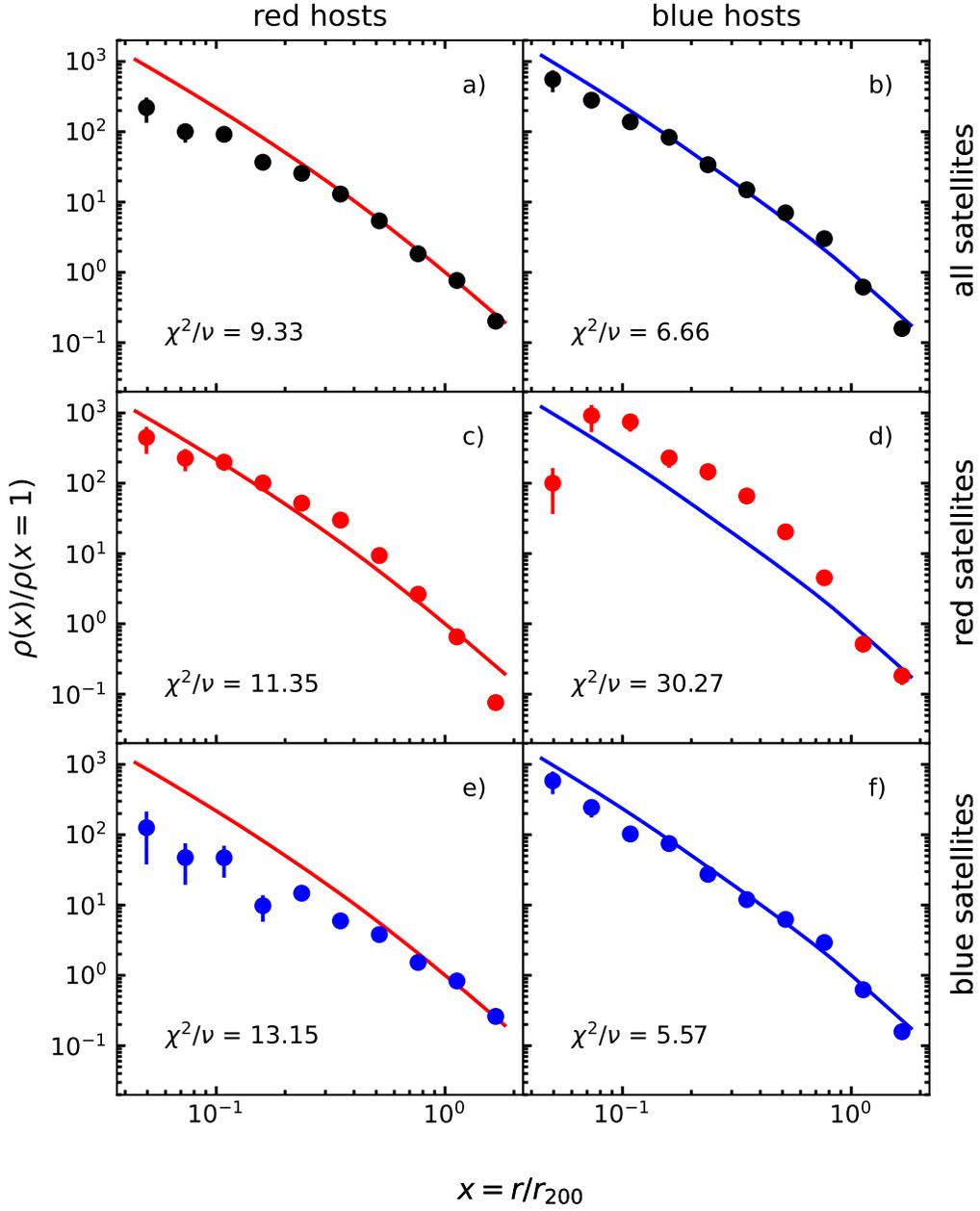}
    \caption{Normalized satellite number density profiles for color-based subdivisions of the
    host-satellite sample.  a) All satellites of red hosts.  b) All satellites of blue hosts.
    c) Red satellites of red hosts.  d) Red satellites of blue hosts.  e) Blue satellites
    of red hosts.  f) Blue satellites of blue hosts.  Solid lines in each panel show the mean mass density of the corresponding
    host galaxies (red line: red hosts; blue line: blue hosts).
    Each panel indicates the $\chi^2$ per degree of freedom for a comparison of the satellite number density
    profile to the mean host mass density profile.}
    \label{red_blue_hosts}
\end{center}
\end{figure}

\begin{figure}
\begin{center}
	\includegraphics[scale=0.95]{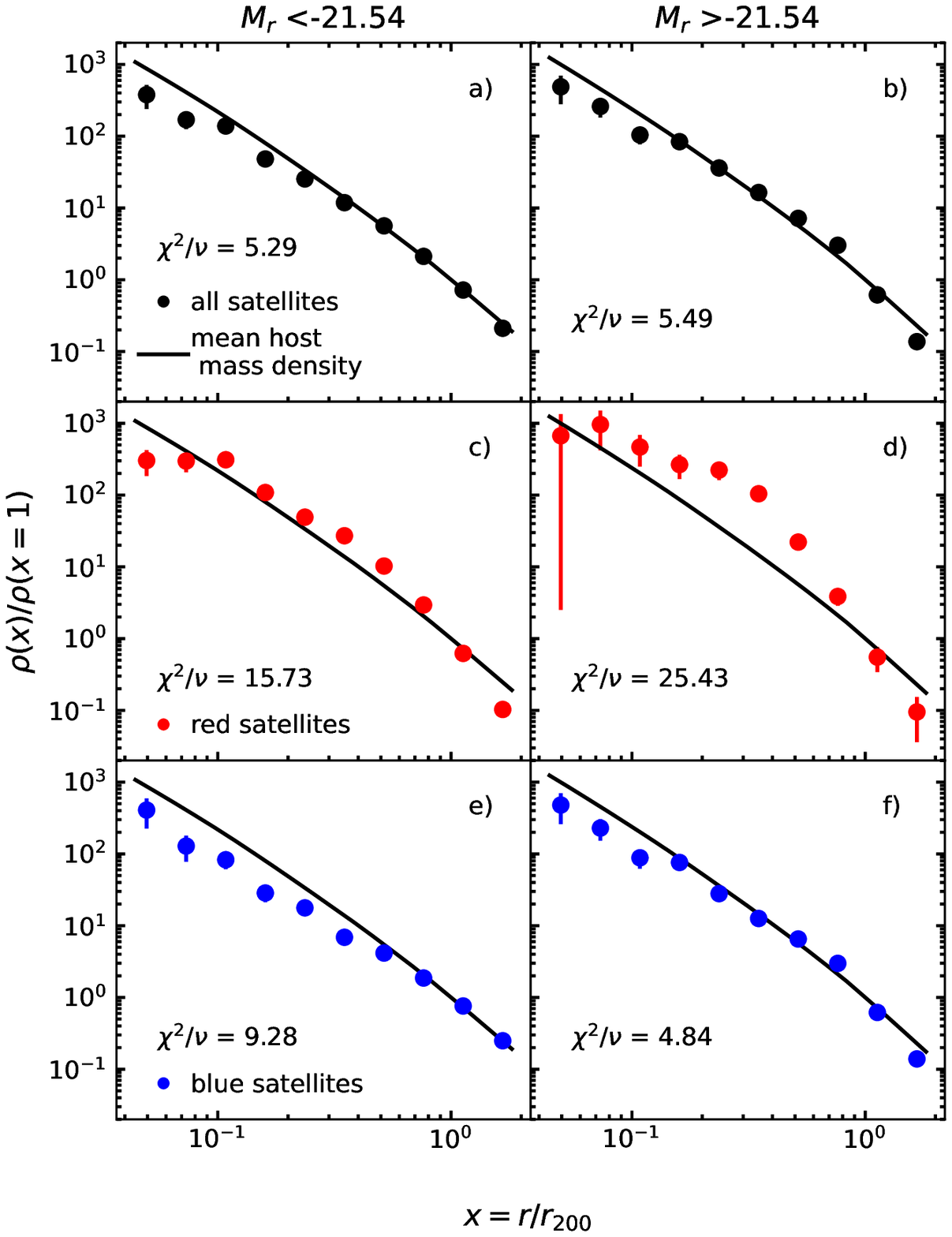}
    \caption{Normalized satellite number density profiles for luminosity-based subdivisions of the
    hosts and color-based subdivisions of the satellites. a) All satellites of the
    brightest 50\% of host galaxies.  b) All satellites of the faintest 50\% of host
    galaxies.  c) Red satellites of the brightest 50\% of host galaxies.  d) Red satellites
    of the faintest 50\% of host galaxies.  e) Blue satellites of the brightest 50\%
    of host galaxies.  f) Blue satellites of the faintest 50\% of host galaxies. 
    Solid black lines in each panel show the the mean mass density of the corresponding
    host galaxies.
    Each panel indicates the $\chi^2$ per degree of freedom for a comparison of the satellite number density
    profile to the mean host mass density profile.}
    \label{bright_faint_hosts}
\end{center}
\end{figure}

\subsection{Dependence of Density Profiles on Host Luminosity and Color}

\begin{figure}
\begin{center}
	\includegraphics[scale=0.95]{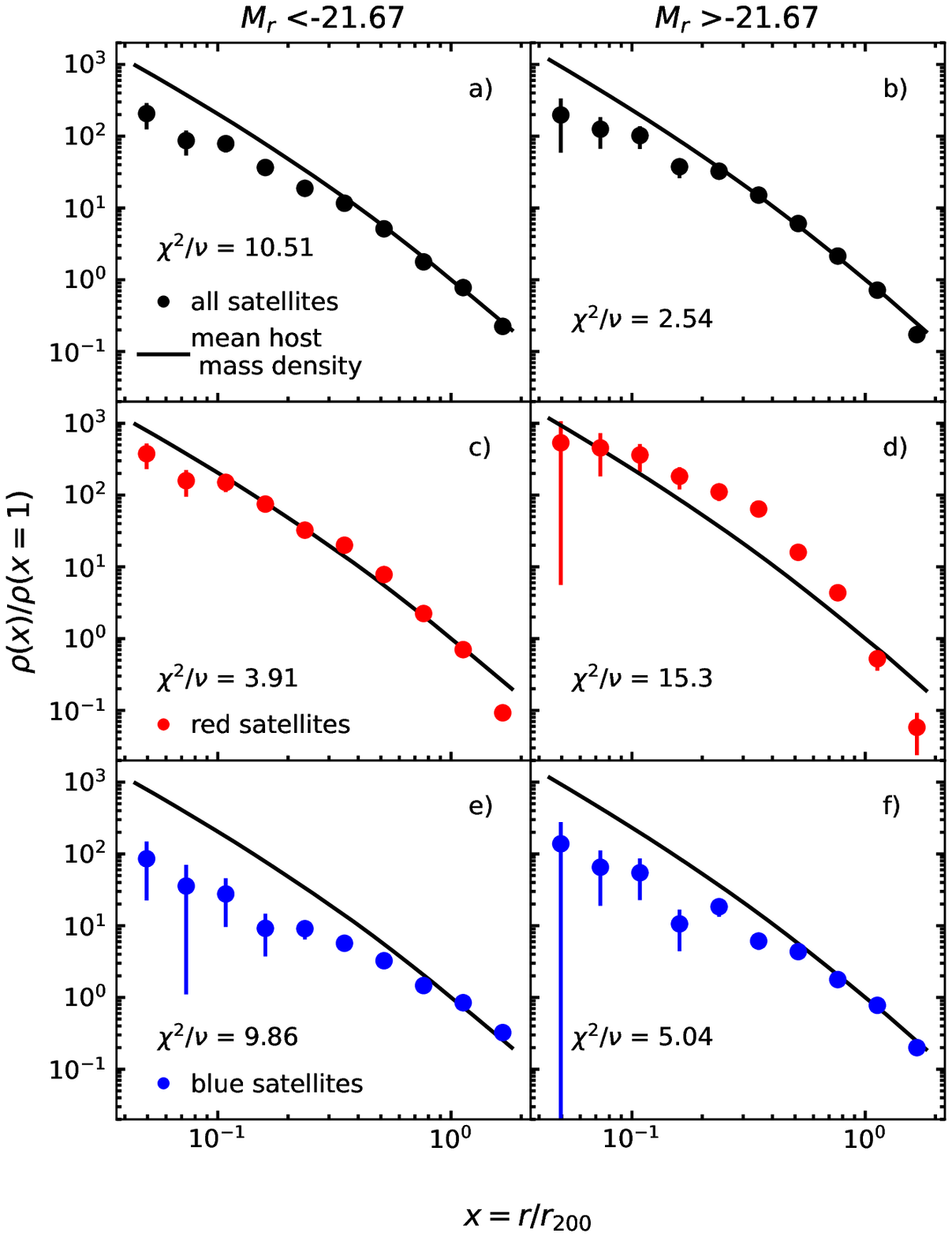}
    \caption{Same as Figure \ref{bright_faint_hosts} but with the sample restricted to red
    host galaxies and their satellites.}
    \label{red_hosts_only}
\end{center}
\end{figure}

\begin{figure}
\begin{center}
	\includegraphics[scale=0.95]{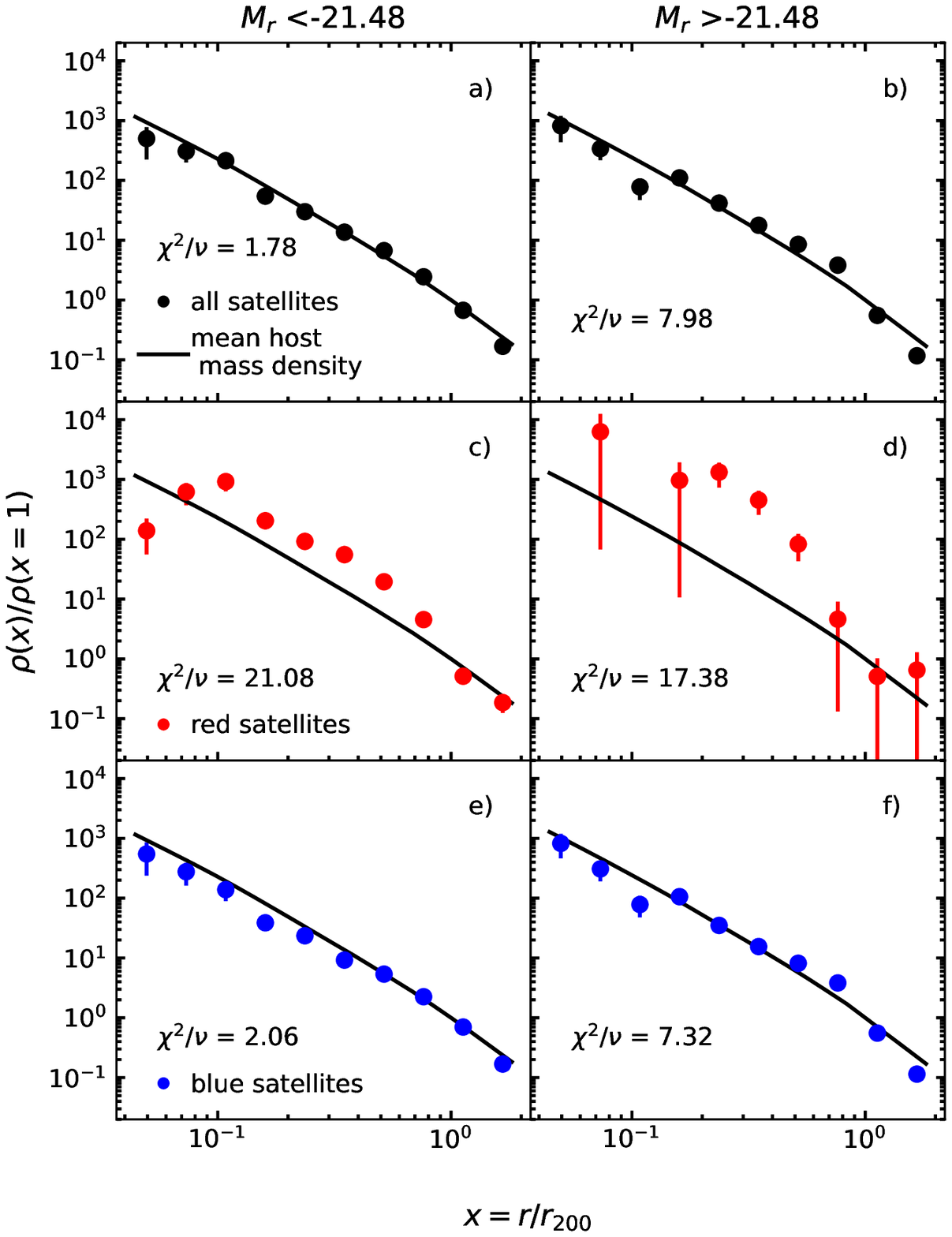}
    \caption{Same as Figure \ref{bright_faint_hosts} but with the sample restricted to blue
    host galaxies and their satellites.}
    \label{blue_hosts_only}
\end{center}
\end{figure}

In Figures \ref{red_blue_hosts}, \ref{bright_faint_hosts}, \ref{red_hosts_only}, and \ref{blue_hosts_only} we subdivide the TNG100 host-satellite sample in various ways in order to explore the degree to which satellites in the subsamples trace the mean mass density profiles of their hosts.  In all of these figures we compare the normalized satellite number density profile to the mean mass density of the corresponding
host galaxies and we list values of the reduced $\chi^2$ 
(i.e., the $\chi^2$ per degree of freedom, $\chi^2 / \nu$) for the comparisons in the individual panels.

Figure \ref{red_blue_hosts} shows results for the host-satellite sample subdivided by host and satellite color.
The left panels of Figure~\ref{red_blue_hosts} show results for the satellites of red host galaxies, while the
right panels show results for the satellites of blue host galaxies.  The top panels of Figure \ref{red_blue_hosts}
show the results for all satellites that surround each type of host galaxy, the middle panels show the results for the red satellites that surround each type of host galaxy, and the bottom panels show the results
for the blue satellites that surround each type of host galaxy.  Formally, none of the host mass density
profiles are a good fit to any of the satellite number density profiles (i.e., $\chi^2 / \nu >> 1$ in all 
cases).  Given that, on average, the red satellites are an older population than are the blue satellites,
it might be expected that the red satellites would be good tracers of the hosts' mass distributions, particularly in the case of red host galaxies.  From Figure \ref{red_blue_hosts}, the red satellites are better tracers of
the mean mass density that surrounds the red hosts than are the blue satellites.  However, the detailed shape of the number density profile for the red satellites of red hosts cannot be fitted by an NFW profile and the value
of the reduced $\chi^2$ (11.35) indicates that red satellites do not trace the mass density of the red hosts well.
The best agreement between a satellite number density profile and a host mass density profile in
Figure \ref{red_blue_hosts} occurs for
the blue satellites of blue hosts (i.e., Figure~\ref{red_blue_hosts}f).  While the agreement is not formally
good for blue satellites of blue hosts ($\chi^2 / \nu = 5.57$), it is better than the agreement for either the red satellites of blue hosts or all satellites of blue hosts. Importantly, the level of disagreement between the host mass distributions and the satellite distributions is such that none of the satellite distributions in Figure \ref{red_blue_hosts} could be used to recover accurate host mass density profiles.

In Figures \ref{bright_faint_hosts}, \ref{red_hosts_only}, and \ref{blue_hosts_only} we explore the degree to which the
satellite distribution traces the mass distribution surrounding hosts of differing luminosity.
In each of these figures, we use the median host magnitude to divide the host sample in half.
In Figure \ref{bright_faint_hosts}, all host galaxies are included in the analyses.  In Figures \ref{red_hosts_only}
and \ref{blue_hosts_only} we further subdivide the hosts by their colors.  As in Figure \ref{red_blue_hosts}, 
in Figures \ref{bright_faint_hosts}, \ref{red_hosts_only} and \ref{blue_hosts_only} we
also subdivide the satellites according to their colors.  Also as in Figure \ref{red_blue_hosts}, none of the host mass density
profiles are a good fit to any of the satellite number density profiles over all host-satellite
separations for which satellites are identified.  In addition, the sense of the disagreement between the host mass density profiles and the number density profiles for the red satellites is often opposite to the
sense of the disagreement between the host mass density and the number
density profiles for the blue satellites.  These systematic, but opposite, differences can combine to give a better agreement between the host mass density profile and the number density profile that is obtained when all satellite
galaxies are included in the calculation than when calculated for either red or blue satellites alone. This effect can be seen in the left columns of Figures \ref{bright_faint_hosts} and \ref{blue_hosts_only} and the right column of Figure \ref{red_hosts_only}.   

Amongst all the results shown in Figures \ref{bright_faint_hosts}, \ref{red_hosts_only}, and \ref{blue_hosts_only}, by 
far the best agreement between the shape of the satellite number density profile and the
shape of the host mass density profile occurs for the complete sample of satellite
galaxies that surround the brightest blue host galaxies (i.e., Figure \ref{blue_hosts_only}a, for
which $\chi^2 / \nu = 1.78$).  This moderate agreement between the spatial distribution
of the complete sample of satellites and the mean mass density surrounding the brightest
blue host galaxies is, nevertheless, somewhat coincidental.  That is, the
number density profile of the blue satellites (which are the majority of the satellites
in this case) falls systematically below the mean mass density of the brightest blue
host galaxies on scales $\lesssim 0.4r_{200}$.  However, the number density profile of the
red satellites rises systematically above the mean mass density of the brightest
blue host galaxies on scales $0.07r_{200} \lesssim r \lesssim 0.7r_{200}$.  It is
only through the combination of these opposite, systematic differences that the shape of the
number density profile of the complete sample of satellite galaxies agrees moderately
well with the shape of the mass density profile of the brightest blue host galaxies.

\subsection{Dependence of Satellite Number Density Profiles on Joining Redshift}

On average, the red TNG100 satellites
joined their hosts at earlier times than did the blue satellites (see \S2).
Because of this, present-day satellite color is a
proxy for the relative ages of the satellites and it might be expected that red satellites will therefore
be the best tracers of the host mass distribution.  However, this expectation presupposes that the earliest satellites to join 
their hosts will, in fact, be good tracers of the host's mass and
we have seen above that red TNG100 satellites are not, in general, good 
tracers of their host's mass distributions.

In an observational sample, the joining times of the satellite galaxies are,
of course, not known. Because of this, it is useful to have a proxy such as 
color to distinguish ``old'' satellites from ``young'' satellites in 
an observational sample.  In a simulated
sample, however, it is straightforward to determine
the redshifts at which the satellites joined their hosts and,
so, it is not absolutely necessary to rely on
proxies to separate ``young'' from ``old'' satellites in a simulated sample.

\begin{figure}
\begin{center}
	\includegraphics[scale=0.95]{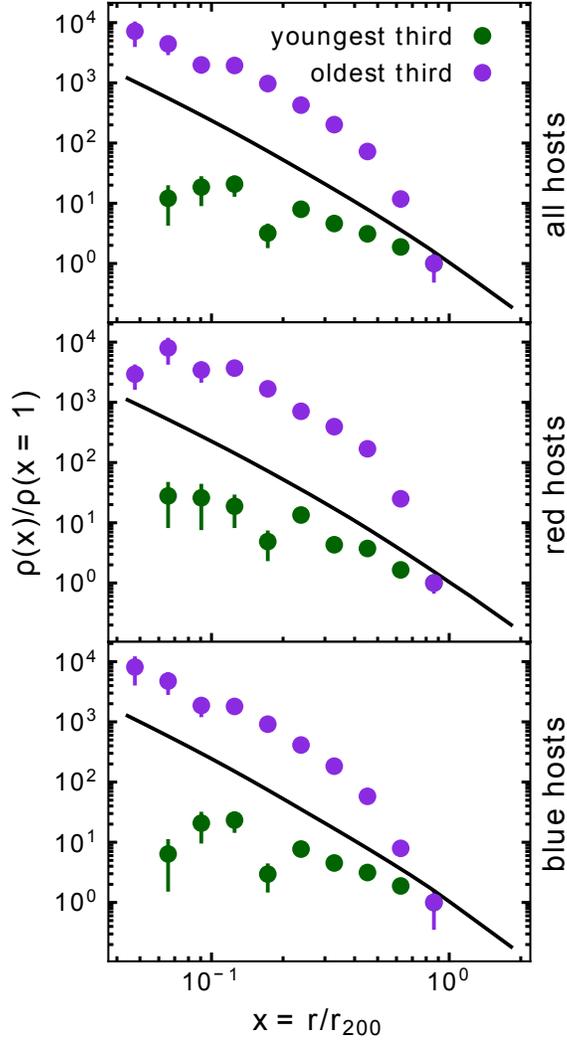}
    \caption{Normalized satellite number density profiles, computed separately for the oldest
    one third of the satellites (purple points) and the youngest one third of the
    satellites (green points). ``Ages'' of the satellites are determined by the
    redshift at which they first entered the virial radius of their host (see text).
    {\it Top:} All host-satellite systems.  
    {\it Middle:} Host-satellite systems with red hosts.
    {\it Bottom:} Host-satellite systems with blue hosts.  In each panel black lines show the
    mean mass densities for the corresponding host
    galaxies.}
    \label{old_young_sats}
\end{center}
\end{figure}

In this section, we compare the normalized number density profiles for subsets of
satellites that joined their hosts in the distant past to
those for satellites that joined their hosts only recently.  To do this,
we divide the full satellite sample into thirds using the joining 
redshift, $z_j$.  Figure \ref{old_young_sats} shows
the normalized number density profiles for the oldest one third of the
satellites (median joining redshift: $z_j^{\rm med} =0.79$) and 
the youngest one third of the satellites (median joining redshift:
$z_j^{\rm med} = 0.074$), along with the mean
host mass density profiles.

From Figure \ref{old_young_sats}, neither the oldest nor the youngest 
satellites have normalized number density profiles that match
the shape of the mean host mass density profiles.  Number
density profiles for the youngest satellites fall systematically
below the host mass density profiles and number density profiles
for the oldest satellites rise systematically above the 
host mass density profiles.  This is true
within the full sample (top panel) as well as samples that are subdivided 
by host color (middle and bottom panels).  

It is unsurprising that the youngest 
satellites do not trace the mass density of their hosts since
the expectation is they have not had time to come into virial
equilibrium with the hosts.  That is, the dynamical time for
a spherical halo is $\tau_{\rm dyn} \sim 0.1 H_0^{-1}$ 
(see, e.g., \citealt{Gan2010}), so
the amount of cosmic time
that passes between $z_j = 0.074$ (i.e., the median joining
redshift of the youngest satellites) and the present is 
$\sim 1$~Gyr, or
$\sim 0.7\tau_{\rm dyn}$.  In the case of the oldest satellites,
the amount of cosmic time that passes between the
median joining redshift ($z_j = 0.79$) and the present
day is $\sim 5\tau_{\rm dyn}$ and, based on the dynamical
time alone, one might expect the oldest satellites to be in 
approximate equilibrium with their hosts' dark matter halos.
However, infalling satellites ought to be subject to dynamical
friction, which will drive them towards the centers of their
hosts' dark matter halos.  

\cite{Gan2010}
showed that the dynamical friction time for satellites is highly
sensitive to the efficiency with which the satellites are
tidally stripped.  However, \cite{Gan2010} also found that
the dynamical friction time for satellites with masses $ < 25$\% of the mass of their host should range from
 $\lesssim 2$~Gyr for satellites on relatively circular orbits
to $\sim 5$~Gyr for satellites on radial orbits, independent of
the degree of tidal stripping.  Since the TNG100 satellites 
are typically orders of magnitude less massive than their 
hosts,
and since the amount of cosmic time
that passes between $z_j = 0.79$ and the present day is 
$\sim 7$~Gyr, it is likely that dynamical friction will 
have affected the present-day locations of the oldest satellites. Therefore,
compared to the mean host mass density, 
the oldest satellites should be more concentrated towards the
centers of the host-satellite systems 
(as, indeed, is in the case in Figure \ref{old_young_sats}).

\section{Summary \& Discussion} \label{disc}

The NFW density profile is motivated by CDM simulations which show that the complex interactions of particles over time cause the mass to settle into this distinct shape, regardless of the cosmology. An important question, then, is whether the spatial
distribution of 
satellite galaxies should be expected to follow a similar profile and, hence, whether
the spatial distribution of satellite galaxies in the observed universe can be
used to directly constrain the concentration of the mass that surrounds the satellites'
host galaxies.

Here we have compared normalized radial number density profiles for the satellites
of isolated host galaxies in the TNG100 simulation to the mean
mass densities of the host galaxies.  The host-satellite systems were selected
using the same criteria that were adopted by \cite{Brainerd2018} for the
Illustris-1 simulation and by \cite{Sales2007} for a galaxy catalog obtained
by applying a SAM to the Millennium simulation.  

In agreement with \cite{Sales2007}, we find that the number density
profile of the complete sample of TNG100 satellites is well-fitted by an 
NFW profile.  In contrast to the results of \cite{Sales2007}, who
concluded that their satellites were only slightly less concentrated
than the host mass distribution, the
concentration of the TNG100 satellites is a factor of $\sim 2$
lower than that of the mean host mass density.  In agreement with 
\cite{Brainerd2018}, we find that the number density profile of
the complete sample of TNG100 satellites does not follow the
shape of the mean host mass density profile; however, despite
being selected in the same way from simulations with identical
initial conditions, there are 
distinct differences between the satellite number density profiles
for the TNG100 and Illustris-1 satellites.  In particular, the distribution of
the TNG100 satellites does not show the steep increase in 
number density at small host-satellite separations that was
found by \cite{Brainerd2018}.  Since the key difference between
the TNG100 simulation and the original Illustris-1 simulation
is the manner in which galactic winds, AGN feedback, and the growth
of supermassive black holes were treated, it is likely that the shape of
the satellite distribution at small host-satellite separations
is particularly sensitive to these these aspects of the galaxy
formation model.

In subdividing the host-satellite sample according to 
host and satellite color, as well as host and satellite
luminosity, we find that no population of TNG100 satellites yields a number
density profile that is formally in good agreement with the shape of the mean
host mass density profile (i.e., $\chi^2 / \nu$ exceeds unity in all cases).  While
there are examples of moderately good agreement when all satellites that surround the
host galaxies are considered, the agreement is somewhat coincidental.  This is due
to the number density profiles of red satellites and blue satellites being
systematically offset in opposite directions
from the mean host mass density profile
(i.e., the number density profile for blue satellites falls below that of the
mass density while the number density profile for red satellites rises above
that of the mass density).  

Unlike \cite{Sales2007}, who found that the distribution of the reddest
satellites was considerably more concentrated than the distribution of
the bluest satellites, we find that, for host-satellite separations
$\lesssim 0.5r_{200}$, the distribution of the reddest
TNG100 satellites is similar to that of the bluest TNG100 satellties.  
The difference between these two results 
may be attributable
to the SAM used by \cite{Sales2007} causing the satellites
to quickly lose their gas reservoirs upon accretion by their hosts.  This would
have stifled future star formation in the accreted satellites, resulting in 
blue satellites turning red over relatively short time scales.

Lastly, if infalling satellites are expected to come into 
equilibrium with their hosts' dark matter halos, 
this should take place over an extended period of
time and satellites that joined their
hosts in the distant past might be expected to be the best
tracers of the mean host mass
density.
However, when we investigate the locations of the satellites as
a function of the redshift at which they joined their hosts, we
find that neither the youngest nor the oldest satellites trace
the mean host mass density. This indicates that, while time since joining 
a system is a major factor in determining the shape of the 
satellite number density profile, it is not
a reliable predictor of the satellite population that best traces
the host mass distribution.  In the case
of the youngest one third of the satellites, not enough time has
passed for their spatial distribution to become as concentrated
as the mean mass density of their hosts.  In the case of the oldest
one third of the satellites, the majority of the satellites 
joined their host sufficiently
far in the past that dynamical friction has likely driven them towards the
centers of the systems, resulting in satellite distributions
that are more concentrated than the mean host mass density.

\section{Conclusions} \label{conc}

Our work shows that the shapes of the TNG100 satellite number density profiles, and the degree to which they agree with their hosts' mass density profiles, are sensitive to the properties of the hosts and satellites that are used in the analysis (i.e., luminosity, color, satellite joining time).  Even samples consisting only of satellites that have had time to virialize within their hosts' halos fail to trace their hosts' mass density. While the oldest satellites are more concentrated than the mean host mass density, the youngest satellites are less concentrated than the mean host mass density. One possible explanation for this phenomenon is that once satellites enter their hosts' halos, dynamical friction drives them towards the centers of the halos. However, host-satellite samples that are drawn from different simulations, but using identical selection criteria, also show distinct differences between their satellite number density profiles (\citealt{Sales2007}, \citealt{Brainerd2018}, this work). This suggests that the satellite number density profiles can also be sensitive to the galaxy formation models adopted by the simulations. The root causes of all of these differences cannot be identified categorically from the work presented here, and we will explore them in more detail in a future analysis. The major conclusion from our analysis is that, if we live in a $\Lambda$CDM universe, it is not clear that the observed satellites of isolated host galaxies will be reliable tracers of their hosts' dark matter halos. Therefore, it is not clear that halo properties of isolated galaxies that are inferred from observed satellite distributions are, in fact, accurate representations of the underlying halo properties.



\begin{acknowledgments}
We are grateful to the anonymous reviewer for helpful suggestions that
improved the paper.
This work was partially supported by National Science 
Foundation grant AST-2009397 and a Summer 2021 Massachusetts 
Space Grant Fellowship.
\end{acknowledgments}

\bibliography{ms_bibliography}{}
\bibliographystyle{aasjournal}



\end{document}